\documentclass[12pt]{article}
\usepackage{graphicx}
\usepackage{here}

\newcommand\pubnumber{SNSN-323-63}
\newcommand\pubdate{\today}

\def\napoli{$^a$Departument of physics, Graduate School of Science and Technology,\\ Tokyo University of Science,\\
2641 Yamazaki, Noda,Chiba 278-0022, Japan\\
$^b$Departument of physics, Faculty of Science and Technology,\\ Tokyo University of Science,\\
2641 Yamazaki, Noda,Chiba 278-0022, Japan\\
$^c$Departument of physics, Graduate School of Science, Kyoto University,\\
Kitashirakawa, Sakyo-ku, Kyoto, 606-8502, Japan\\
$^d$Departument of Applied Physics, Faculty of Engineering, University of Miyazaki\\
1-1 Gakuen Kibana-dai Nishi, Miyazaki 889-2192, Japan}

\def\Title#1{\begin{center} {\Large #1 } \end{center}}
\def\Author#1{\begin{center}{ \sc #1} \end{center}}
\def\Address#1{\begin{center}{ \it #1} \end{center}}

\newcommand\pubblock{\rightline{\begin{tabular}{l} \pubnumber\\
         \pubdate  \end{tabular}}}
\newenvironment{Abstract}{\begin{quotation}  }{\end{quotation}}
\newenvironment{Presented}{\begin{quotation} \begin{center} 
             PRESENTED AT\end{center}\bigskip 
      \begin{center}\begin{large}}{\end{large}\end{center} \end{quotation}}
\def\Acknowledgements{\bigskip  \bigskip \begin{center} \begin{large}
             \bf ACKNOWLEDGEMENTS \end{large}\end{center}}

\def\beq{\begin{equation}}
\def\eeq#1{\label{#1}\end{equation}}
\def\eeqn{\end{equation}}

\def\beqa{\begin{eqnarray}}
\def\eeqa#1{\label{#1}\end{eqnarray}}
\def\eeqan{\end{eqnarray}}

\let\bar=\overbar

\def\Dslash{\not{\hbox{\kern-4pt $D$}}}
\def\dslash{\not{\hbox{\kern-2pt $\del$}}}

\def\msb{{\bar{\ssstyle M \kern -1pt S}}}

\begin{document}
\begin{titlepage}
\pubblock

\vfill
\Title{Study of the basic performance of the XRPIX for the future astronomical X-ray satellite}
\vfill
\Author{Koki Tamasawa$^a$, Takayoshi Kohmura$^a$, Takahiro Konno$^b$\\
Takeshi Go Tsuru$^c$, Takaaki Tanaka$^c$,Ayaki Takeda$^c$,\\ Hideaki Matsumura$^c$,
Koji Mori$^d$, Yusuke Nishioka$^d$,\\ and Ryota Takenaka$^d$
}
\Address{\napoli}
\vfill
\begin{Abstract}
We have developed CMOS imaging sensor (XRPIX) using SOI (Silicon-On-Insulator) technology for the X-ray astronomical use. 
XRPIX(X-Ray soiPIXel) has advantage of a high time resolution, a high position resolution and an observation in a wide X-ray energy band with a thick depletion layer of over 200$\mu$m.
However, the energy resolution of XRPIX is not as good as one of X-ray CCD. Therefore improvement of the the energy resolution is one of the most important development item of XRPIX.\par
In order to evaluate the performance XRPIX more precisely, we have investigated on the temperature dependence of the basic performance, such as readout noise, leak current, gain and energy resolution, using two type of XRPIX, XRPIX1 and XRPIX2b\_CZ. \par
In our study, we confirmed the readout noise, the leak current noise and the energy resolution clearly depended on the operating temperature of XRPIX. 
In addition, we divided the readout noise into the leak current noise and the circuit origin noise.
As a result, we found that noise of the electronic circuitry origin was  proportional to the square root of operating temperature.
\end{Abstract}
\vfill
\begin{Presented}
International Workshop on SOI Pixel Detector (SOIPIX2015), Tohoku University, Sendai, Japan, 3-6, June, 2015.
\end{Presented}
\vfill
\end{titlepage}
\def\thefootnote{\fnsymbol{footnote}}
\setcounter{footnote}{0}

\section{Introduction}
The X-ray CCD has been the standard imaging spectrometer with low noise, high position sensitivity and moderate energy resolution on board X-ray astronomical satellite, 
and the X-ray CCD  has played an important rule to observe all X-ray objects in space. 
However, due to pile up effect resulting from its low time resolution of a few second, the X-ray CCD is not suitable for the study on the high time variation in X-ray which are shown in black hole binaries, isolated neutron stars, and so on.\par
Therefore, we have developed XRPIX using SOI as a next generation X-ray detector equivqlent performance both in the position sensitivity and in the energy resolution to the X-ray CCD.
The XRPIX has implemented the correlated double sampling (CDS) function in each pixel to reduce the noise and has also implemented the event trigger output function to readout only X-ray event pixel which enable the higher time resolution of a few $\mu$ sec. This higher time resolution of XRPIX with event trigger out put function enables the anti coincidence measurement in order to remove the NXB (Non X-ray Background).  
The XRPIX has a thick depletion layer of over 200$\mu$m which enables to cover from the soft X-ray band below 1keV and the high X-ray energy band above a few tens of keV.
~\cite{Ryu_eng}~\cite{takeda_1}.

\section{XRPIX of detail}
\subsection{XRPIX1}
We have developed a new type semiconductor pixel sensor referred to as "XRPIX" on the basis of SOI CMOS technology for X-ray astronomical use. 
This SOI pixel sensor is a laminated structure using a low-resistivity Si bonded wafer for high-speed CMOS circuits, an SiO2 insulator, and a high-resistivity depleted Si layer for X-ray detection. 
The CMOS circuit implemented in each pixels has the CDS function and the event trigger output function. 
Thanks to these two functions, XRPIX enables to cancel the reset noise in the CMOS circuit and to read out analog signal at a high speed~\cite{Ryu_eng}.\par
XRPIX1(X-ray soiPixel 1st) is the first XRPIX series.
It is 2.4$\times$2.4 mm$^2$ in size and consists of 32$\times$32 pixels. The pixel size of 30.6$\times$30.6 $\mu$m$^2$, 
so the effective sensing area of the approximately 980$\times$980  $\mu$m$^2$.
In addition, It include a thick high-resistivity Si-sensor layer (thickness 260$\mu$m, resisrivity of 700$\Omega$$\cdot$cm).\par
\begin{figure}[H]
\centering
\includegraphics[height=1.8in]{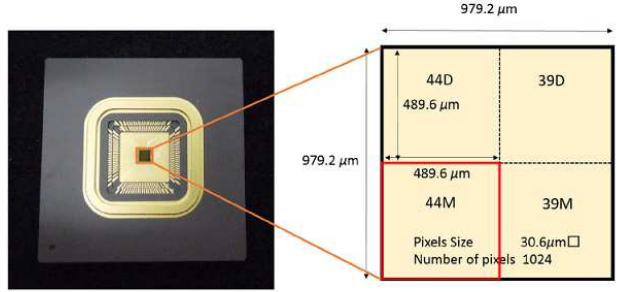}
\caption{(Left) package photo of XRPIX1.
 (Right) block diagram of the X-ray imaging area.
 The evaluation of the area enclosed by a red line.}
\label{fig:XRPIX1}
\end{figure}

For evaluation, XRPIX1 is separated into 4 blocks according to the different types of transistors and capacitors 
used in the pixel circuitry (Figure \ref{fig:XRPIX1} right).
In this paper, we focus on the block consisting of the source-tie and the body-tie type transistors suitable for use in low temperature.

\subsection{XRPIX2b\_CZ}
XRPIX2b, compared to XRPIX1, is expand the X-ray imaging area by increase in the number of pixels 
and the energy resolving, noise and gain improved.
In addition, XRPIX2b can evnt driven readout stably~\cite{takeda_2}.\par
XRPIX2b exist two types (CZ-type and FZ-type) by difference of thick high-resistivity Si-sensor layer.
They are  6.0$\times$6.0 mm$^2$ in size and consists of 144$\times$144 pixels with the pixel size of 30$\times$30 $\mu$m$^2$, 
so the effective sensing area of the approximately 4.3$\times$4.3 mm$^2$.
In addition, It include a thick high-resistivity Si-sensor layer (CZ-type and FZ-tipe are thickness 260$\mu$m and 500$\mu$m and resisrivity of 1.5k$\Omega$$\cdot$cm and 5k$\Omega$$\cdot$cm, respectively).

\section{Experiment Description and Evaluation Results}
\subsection{Experiment Setup}
\begin{figure}[H]
\centering
\includegraphics[height=2.5in]{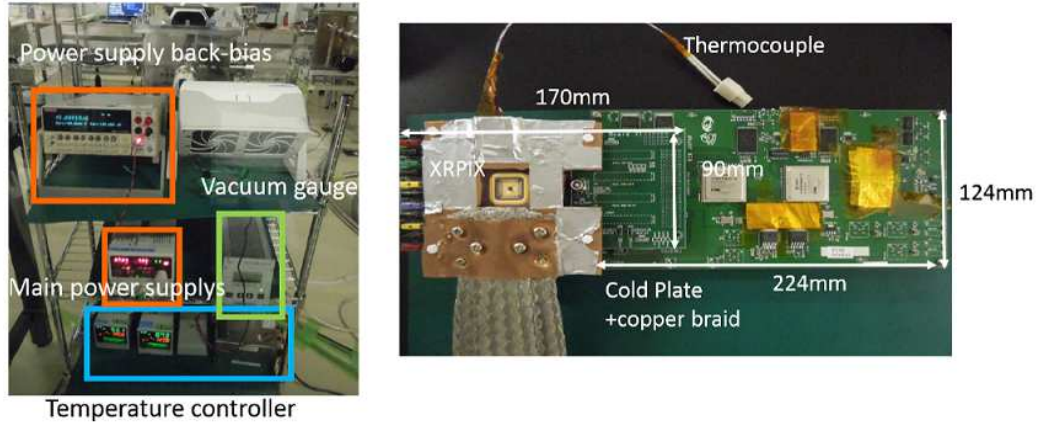}
\caption{(Left) Experimental system.
(Right) XRPIX1 and readout board.
Cold plate is provided for cooling the XRPIX. The temperature of the cold plate is equal to the temperature of the XRPIX. 
The vacuum level and XRPIX temperature was 10$^{-6}$ hPa and from 20 to -70 ℃, respectively.}
\label{fig:SEABAS}
\end{figure}

In order to investigate the temperature dependence of the XRPIX basic performance, 
while changing the temperature, we measured the output of the chip under the condition of irradiation and not. 
Figure \ref{fig:SEABAS} shows the experimental system.
In the experiments, the operating temperature can be controlled in the range 20 to -70℃.

\subsection{XRPIX1}
\subsubsection{X-ray Responsivity}
\begin{figure}[H]
\centering
\includegraphics[height=1.8in]{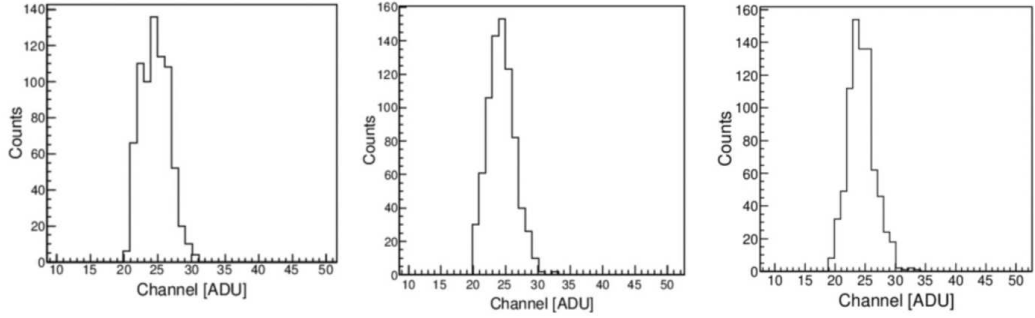}
\caption{Spectrum obtained by XRPIX1 irradiated by the Mn-K$\alpha$ X-ray, $^{55}$Fe. (Left) The operating temperature is -17℃. (Center) The operating temperature is -30℃. (Right) The operating temperature is -50℃.}
\label{fig:XRPIX1 55Fe}
\end{figure}

It was examined the response of the chip by irradiating X-rays from the $^{55}$Fe radio isotope.
Table \ref{tab:XRPIX1 55Fe cond} and Figure \ref{fig:XRPIX1 55Fe} show experiment condition and the spectrum of the X-ray emission from a $^{55}$Fe radio isotope, respectively.

\begin{table}[H]
\begin{center}
\caption{X-ray, from the $^{55}$Fe radio isotope, irradiation experiment condition.}
\begin{tabular}{cc} \hline \hline
Temperature& 22, -17, -30, -50\ [℃]\\ 
Back bias& 5\ [V]\\
The number of getting frame& 100$\times$10$^5$\ [frame]\\
Integration Time& 1.0\ [ms]\\ \hline
\end{tabular}
\label{tab:XRPIX1 55Fe cond}
\end{center}
\end{table}

Figure \ref{fig:XRPIX1 55Fe} shows spectrum overlap Mn-K$\alpha$ and Mn-K$\beta$.
In order to determine the central value, the width and the height of the spectrum, 
the spectrum was approximated by one of the Gaussian function and fitting. 
It was calculated gain and energy resolution of the calculated central value and wide.
The gain G [$\mu$ V/e$^-$] is expressed as,
\begin{equation}
G = \frac{\mathrm{central\ value}}{5.895} \times 244 \times W_{\mathrm{Si}} 
\label{eq:gain}
\end{equation}
The energy resolution FWHM [keV]  is expressed as,
\begin{equation}
\mathrm{FWHM}= 2\sqrt{2\mathrm{ln}2} \times \mathrm{wide} \div G \times 244 \times W_{\mathrm{Si}} 
\label{eq:Erev}
\end{equation}
1 ADU is 244 [$\mu$V]. \textit{W$_{Si}$} is the average ionization energy of Si and is 3.65 [eV/e$^-$].

\begin{figure}[H]
 \begin{minipage}{0.5\hsize}
  \begin{center}
\includegraphics[height=2.5in,angle=270]{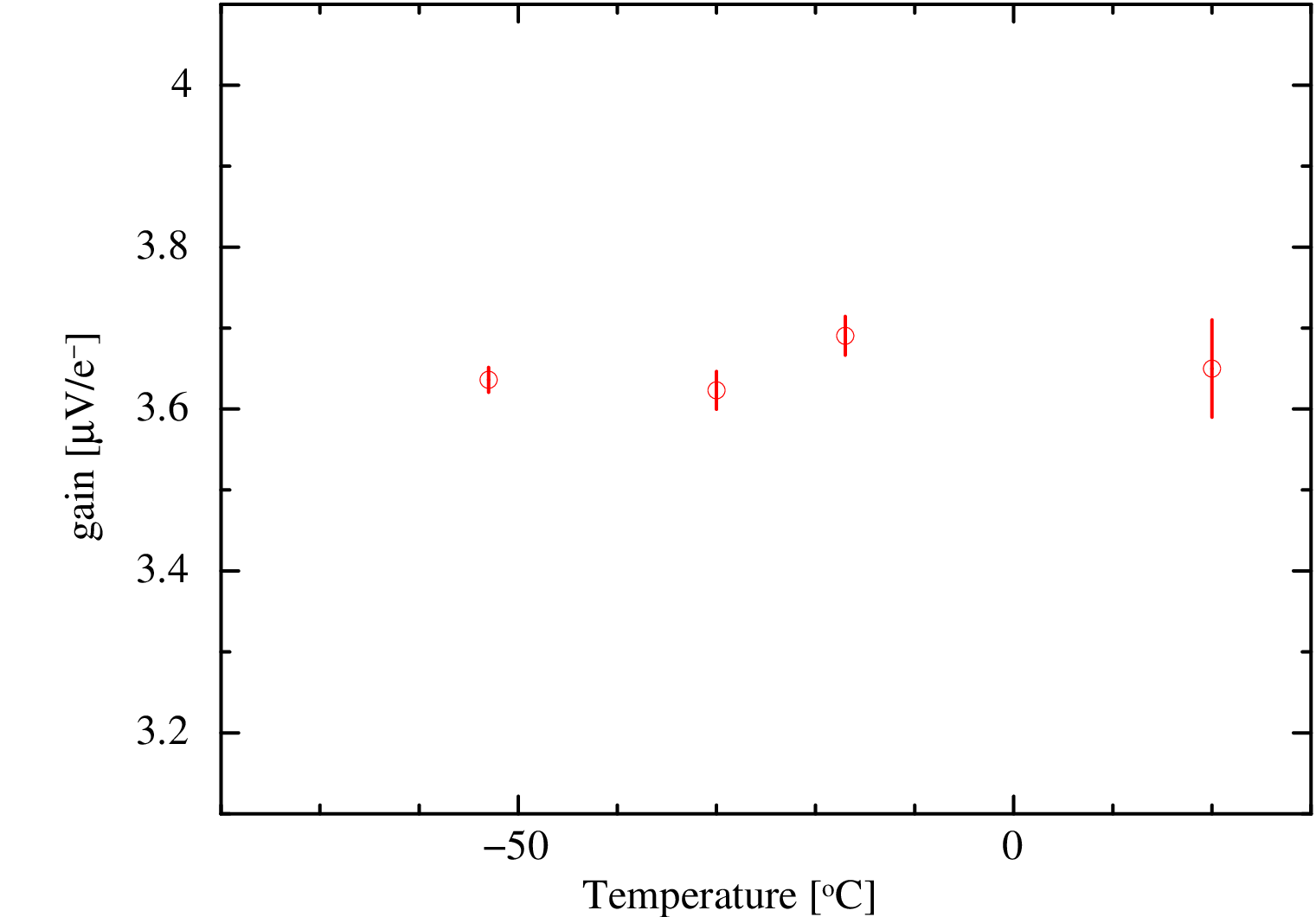}
\caption{Temperature dependence of the gain of XRPIX1.}
\label{fig:XRPIX1 gain}
\end{center}
 \end{minipage}
 \begin{minipage}{0.5\hsize}
  \begin{center}
 \includegraphics[height=2.4in,angle=270]{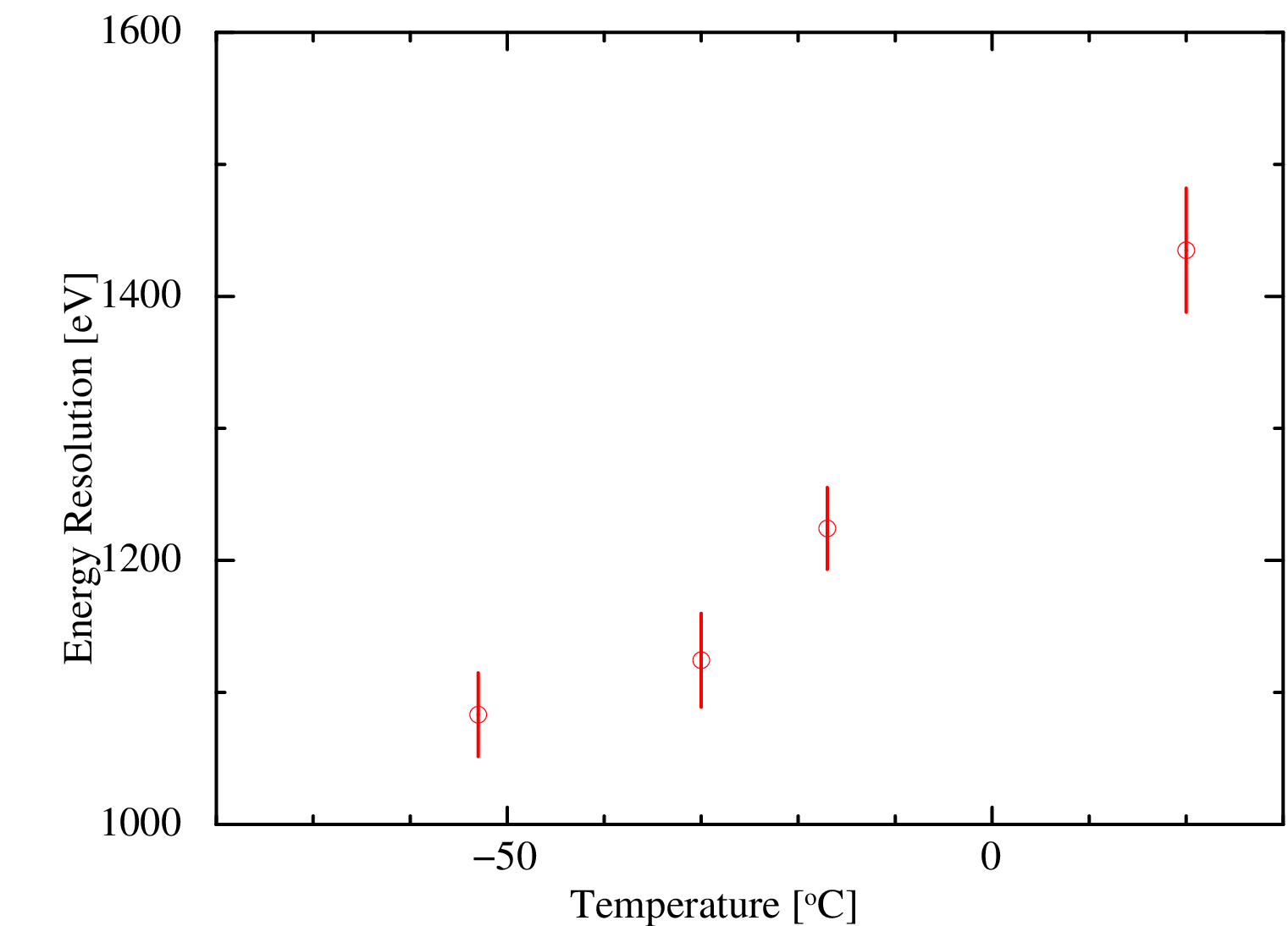}
\caption{Temperature dependence of the energy resolution of XRPIX1 (FWHM)of Mn-Kα of 55Fe radio isotope.}
\label{fig:XRPIX1 Erev}
\end{center}
 \end{minipage}
\end{figure}

Figure \ref{fig:XRPIX1 gain} and Figure \ref{fig:XRPIX1 Erev} show temperature dependence of the gain 
and temperature dependence of the energy resolution (FWHM) of Mn-K$\alpha$ of $^{55}$Fe radio isotope, respectively.



In Figure \ref{fig:XRPIX1 gain}, gain does not depend on operating temperature. Hereafter gain is 3.65 [$\mu$V/e$^-$].
In Figure \ref{fig:XRPIX1 Erev}, Energy resolution is reduced exponentially as the temperature decreases.

\subsubsection{About noise}
\begin{figure}[H]
 \begin{minipage}{0.5\hsize}
  \begin{center}
\includegraphics[height=2.5in,angle=270]{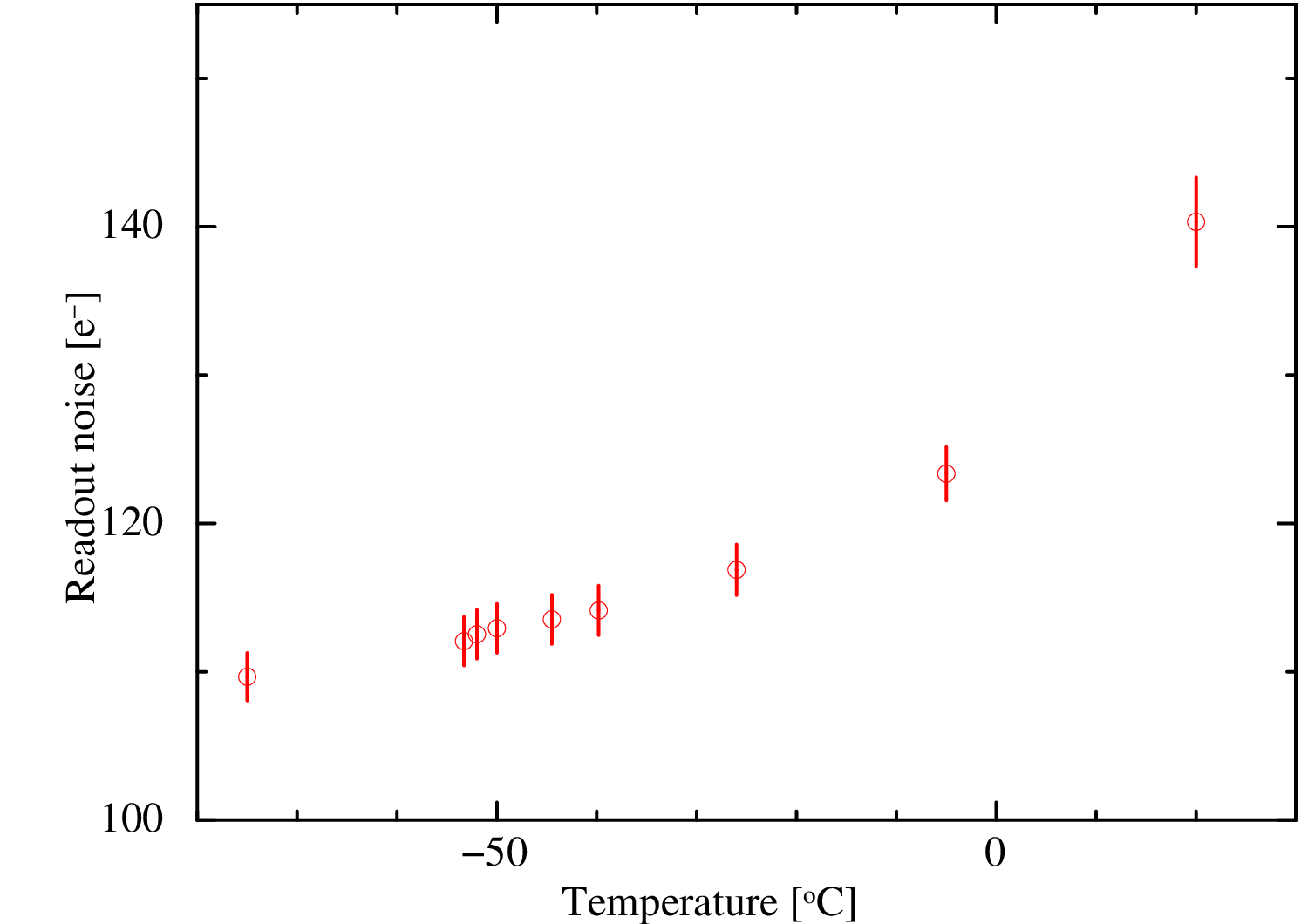}
\caption{Temperature dependence of the readout noise of XRPIX1.}
\label{fig:XRPIX1 readout}
\end{center}
 \end{minipage}
 \begin{minipage}{0.5\hsize}
  \begin{center}
 \includegraphics[height=2.5in,angle=270]{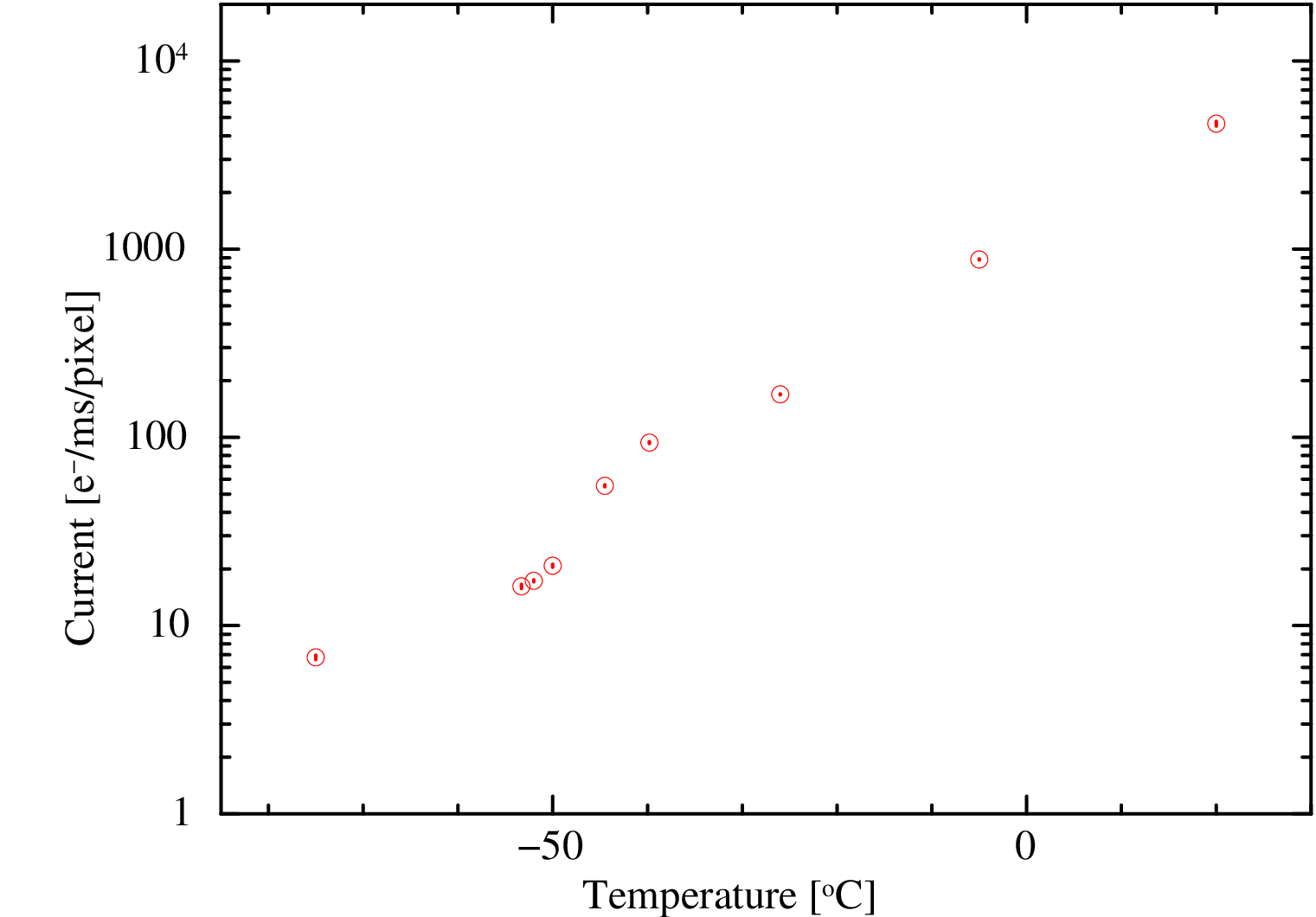}
\caption{Temperature dependence of the leak current of XRPIX1.}
\label{fig:XRPIX1 leak}
\end{center}
 \end{minipage}
\end{figure}

We show the results of noise measurements by changing operating temperature(20,-5.-26.-40,-45,-50,-75℃).
The measument noise values are converted from ADU into electrons with the gain.
Figure \ref{fig:XRPIX1 readout} and Figure \ref{fig:XRPIX1 leak} show temperature dependence of the readout noise 
and temperature dependence of the leak current, respectively.

In Figure \ref{fig:XRPIX1 readout}, it was found that readout noise is reduced as the operating temperature decreases.
In Figure \ref{fig:XRPIX1 leak}, it was found that leak current is reduced exponentially as the operating temperature decreases.

\subsection{XRPIX2b\_CZ}
\subsubsection{X-ray Responsivity}
\begin{figure}[H]
\centering
\includegraphics[height=1.6in]{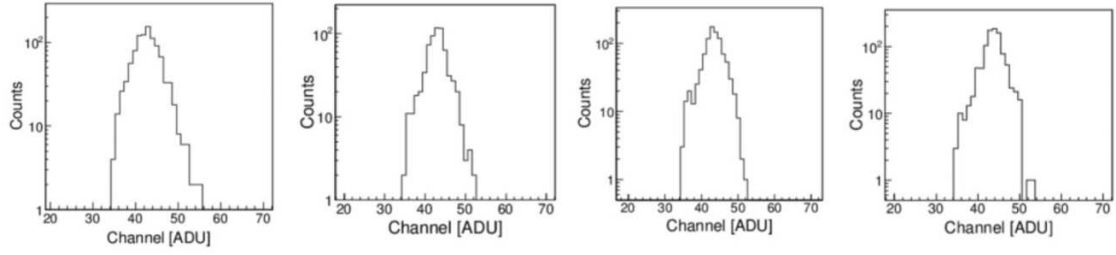}
\caption{Spectrum obtained by XRPIX2b\_CZ irradiated by the Mn-K$\alpha$ X-ray, $^{55}$Fe. The operating temperature is 22, -15, -43, -68℃.}
\label{fig:XRPIX2b 55Fe}
\end{figure}

It was examined the response of the device by irradiating X-rays from the $^{55}$Fe radio isotope.
Table \ref{tab:XRPIX2b 55Fe cond} and Figure \ref{fig:XRPIX2b 55Fe} show experiment condition and the spectrum of the X-ray emission from a $^{55}$Fe radio isotope, respectively.

\begin{table}[H]
\begin{center}
\caption{X-ray, from the $^{55}$Fe radio isotope, irradiation experiment condition.}
\begin{tabular}{cc} \hline \hline
Temperature& 22, -15, -43, -68\ [℃]\\ 
Back bias& 5\ [V]\\
The number of getting frame& 20$\times$10$^5$\ [frame]\\
Integration Time& 0.1\ [ms]\\ \hline
\end{tabular}
\label{tab:XRPIX2b 55Fe cond}
\end{center}
\end{table}

As with XRPIX1, spectrum overlap Mn-K$\alpha$ and Mn-K$\beta$.
The spectrum was approximated by one of the Gaussian function and ftting. 
It was calculated gain and energy resolution by eq(\ref{eq:gain}) and eq(\ref{eq:Erev}).

\begin{figure}[H]
 \begin{minipage}{0.5\hsize}
  \begin{center}
\includegraphics[height=2.5in,angle=270]{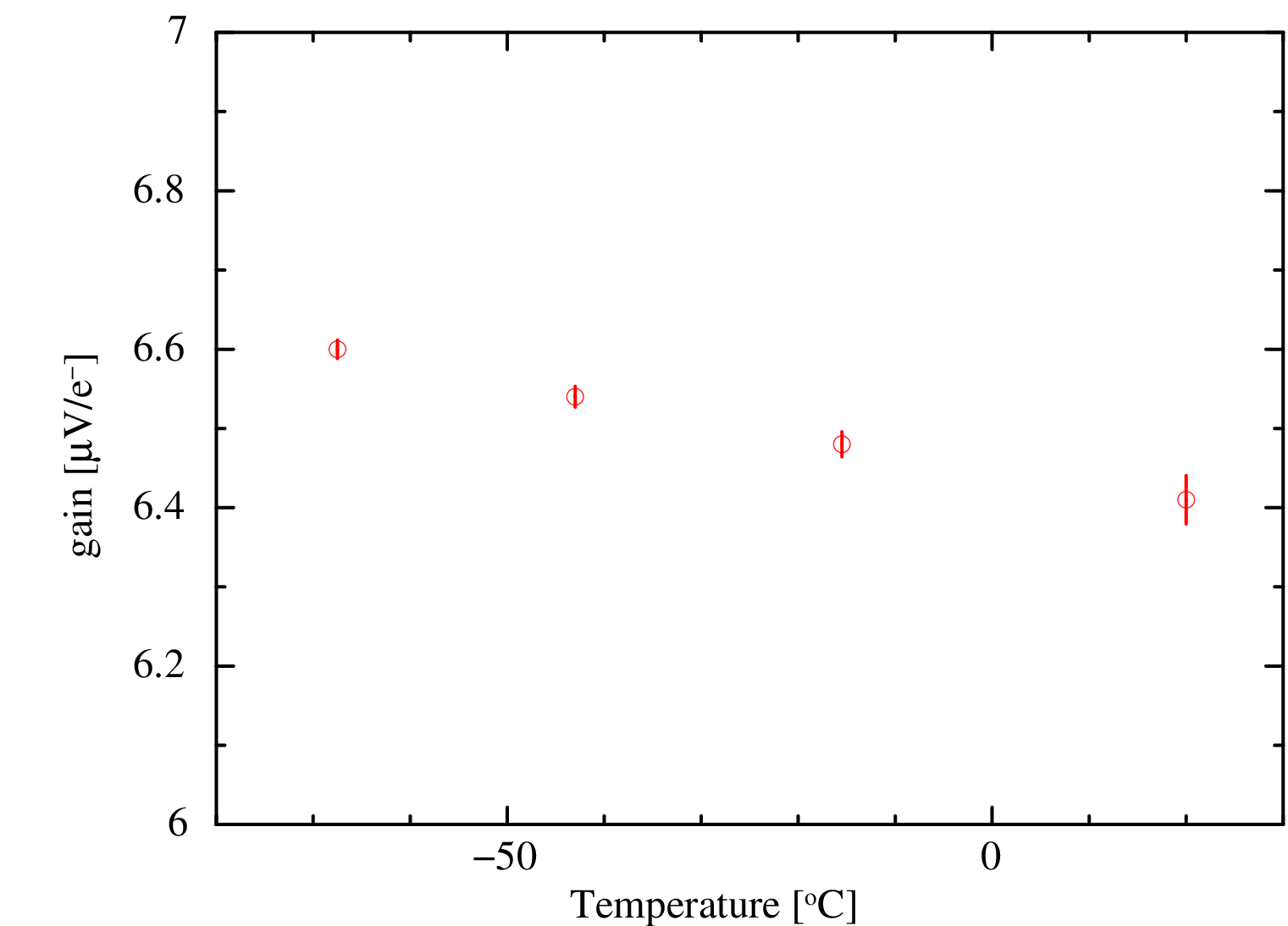}
\caption{Temperature dependence of the gain of XRPIX2b\_CZ.}
\label{fig:XRPIX2b gain}
\end{center}
 \end{minipage}
 \begin{minipage}{0.5\hsize}
  \begin{center}
\includegraphics[height=2.4in,angle=270]{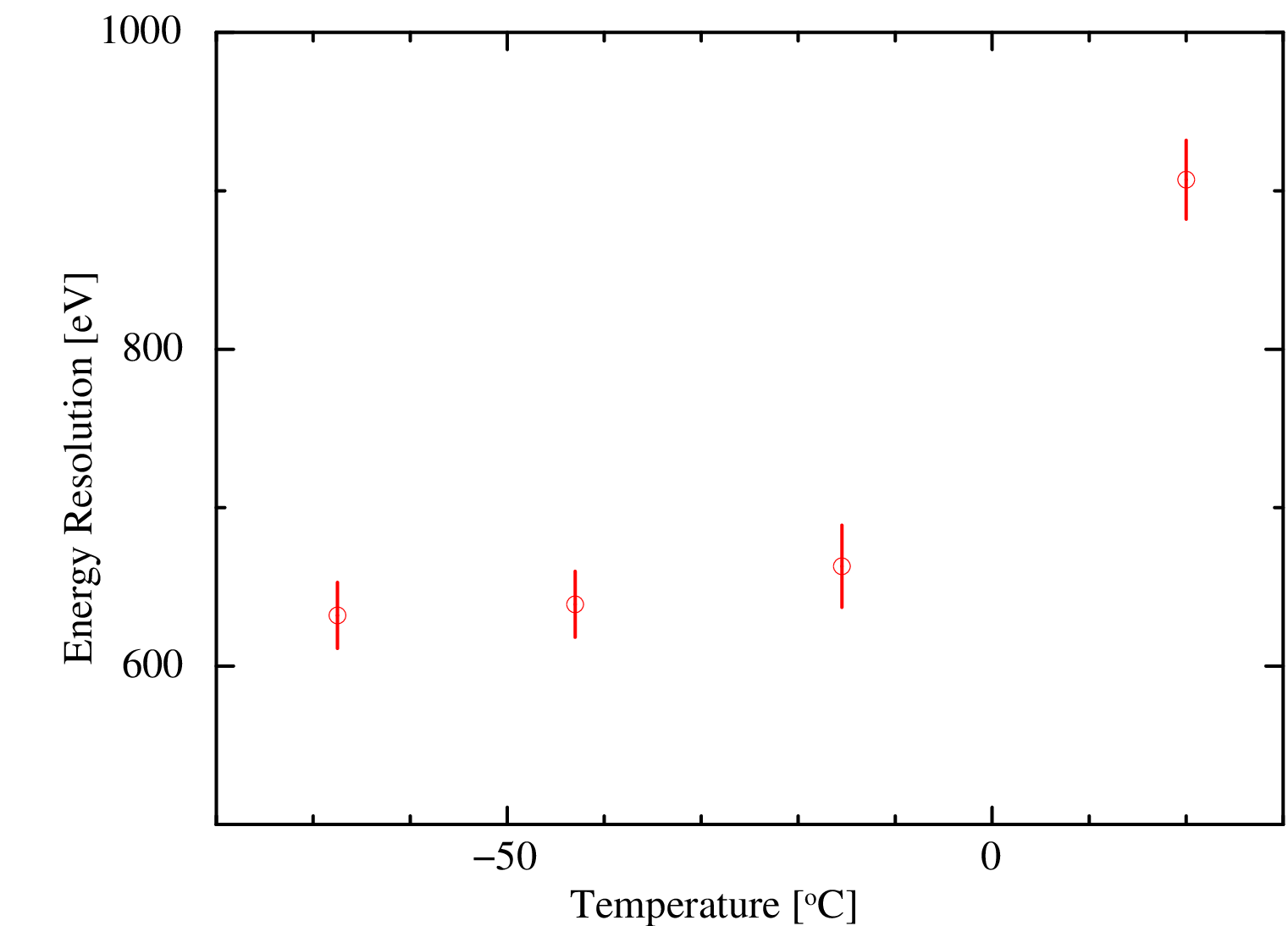}
\caption{Temperature dependence of the energy resolution of XRPIX2b\_CZ (FWHM)of Mn-Kα of 55Fe radio isotope.}
\label{fig:XRPIX2b Erev}
\end{center}
 \end{minipage}
\end{figure}

Figure \ref{fig:XRPIX2b gain} and Figure \ref{fig:XRPIX2b Erev} show temperature dependence of the gain 
and temperature dependence of the energy resolution (FWHM)of Mn-K$\alpha$ of $^{55}$Fe radio isotope, respectively.

In Figure \ref{fig:XRPIX2b gain},unlike XRPIX1, it was found that gain depend on operating temperature.
In Figure \ref{fig:XRPIX2b Erev}, it was found that Energy resolution is reduced exponentially as the temperature decreases. 
Compared to XRPIX1, gain and Energy resolution of XRPIX2b\_CZ became about 2 times and about 0.5 times, respectively.

\subsubsection{About noise}
\begin{figure}[H]
 \begin{minipage}{0.5\hsize}
  \begin{center}
\includegraphics[height=2.5in,angle=270]{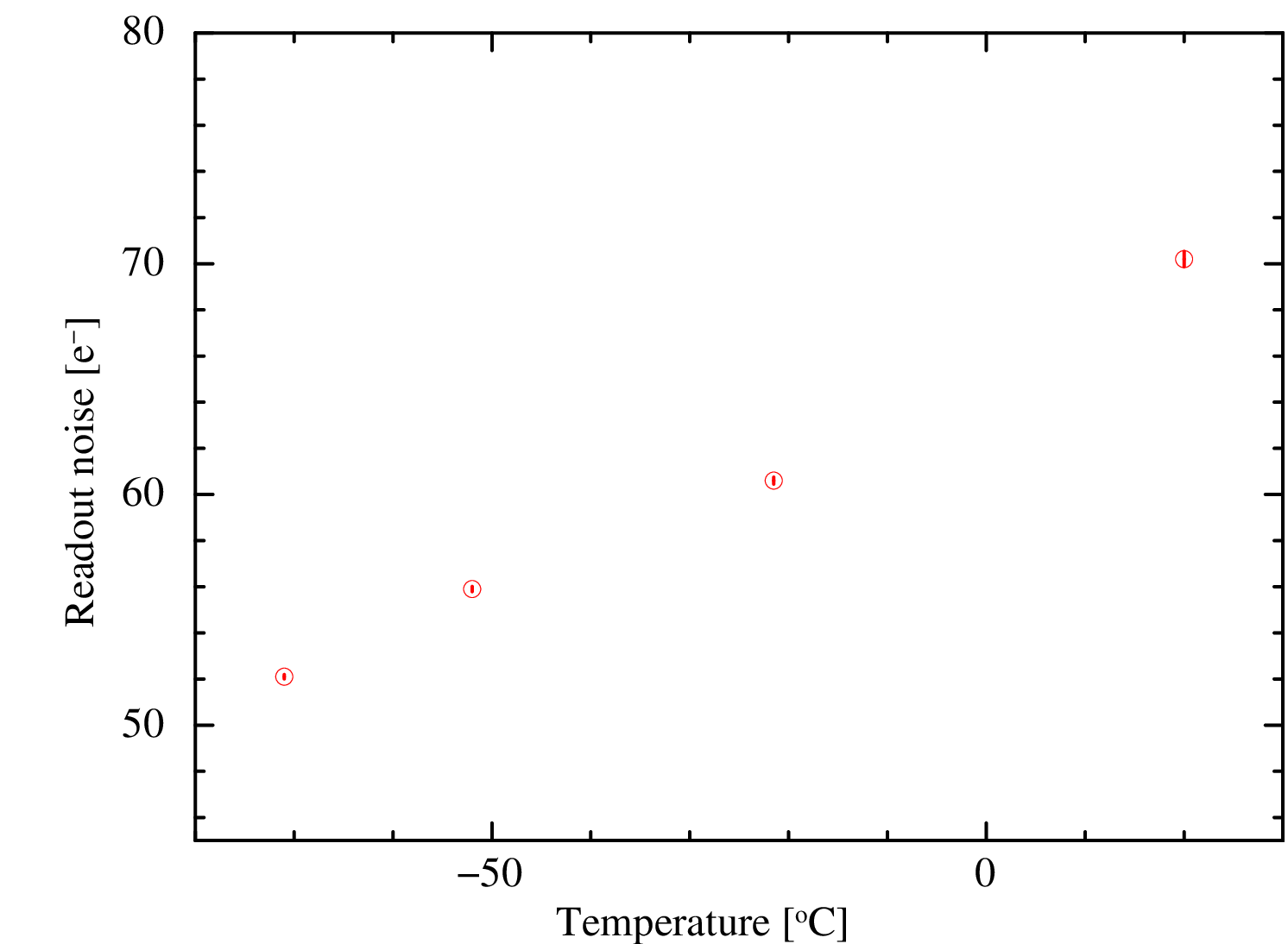}
\caption{Temperature dependence of the readout noise of XRPIX2b.}
\label{fig:XRPIX2b readout}
\end{center}
 \end{minipage}
 \begin{minipage}{0.5\hsize}
  \begin{center}
\includegraphics[height=2.5in,angle=270]{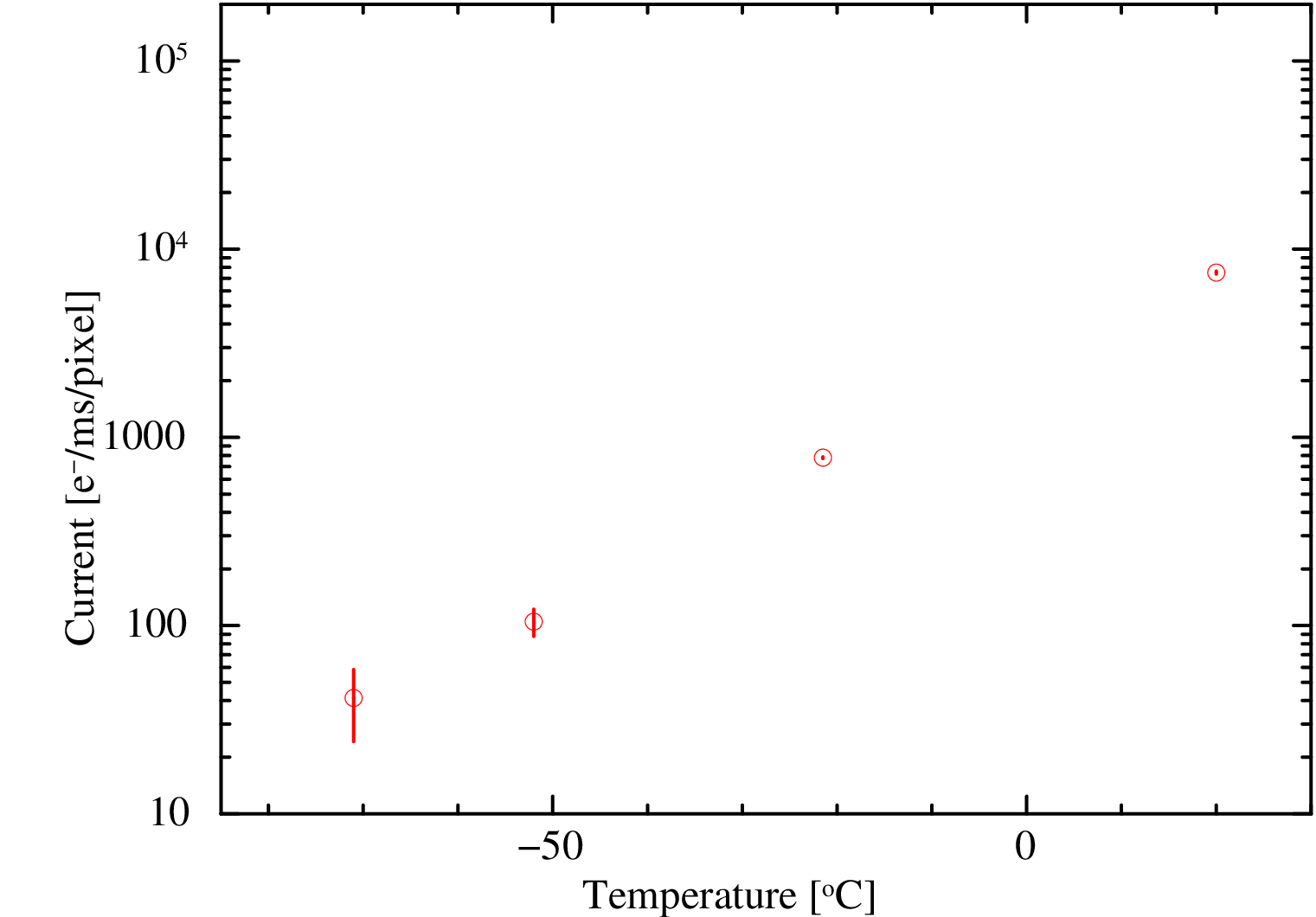}
\caption{Temperature dependence of the leak current of XRPIX2b.}
\label{fig:XRPIX2b leak}
\end{center}
 \end{minipage}
\end{figure}

We show the results of noise measurements by changing operating temperature(20,-20,-52,-71℃)
Figure \ref{fig:XRPIX2b readout} and Figure \ref{fig:XRPIX2b leak} show temperature dependence of the readout noise 
and temperature dependence of the leak current, respectively.

In Figure \ref{fig:XRPIX2b readout}, it was found that readout noise is reduced as the operating temperature decreases.
In Figure \ref{fig:XRPIX2b leak}, it was found that leak current is reduced exponentially as the operating temperature decreases.
Compared to XRPIX1, readout noise and leak current of XRPIX2b\_CZ became about 0.5 times and about 10 times (wrong), respectively.

\subsection{Evaluation of the circuit origin noise}
We show the evaluation reslut of the circuit origin noise.
If readout noise \textit{N} is assumed to consist of the circuit origin noise cicult and the leak current \textit{$\sigma^2_I$} origin noise, the readout noise is expressed as,
\begin{equation}
N = \sqrt{\sigma^2_\mathrm{{cicult}} + \sigma^2_I \times \mathrm{Exposure Time}}
\label{eq:N}
\end{equation}
Moreover, (\ref{eq:cicult}) is obtained from (\ref{eq:N}).
\begin{equation}
\sigma^2_\mathrm{{cicult}} = N^2 - \sigma^2_I \times \mathrm{Exposure Time}
\label{eq:cicult}
\end{equation}

\begin{figure}[H]
\centering
\includegraphics[height=2.5in]{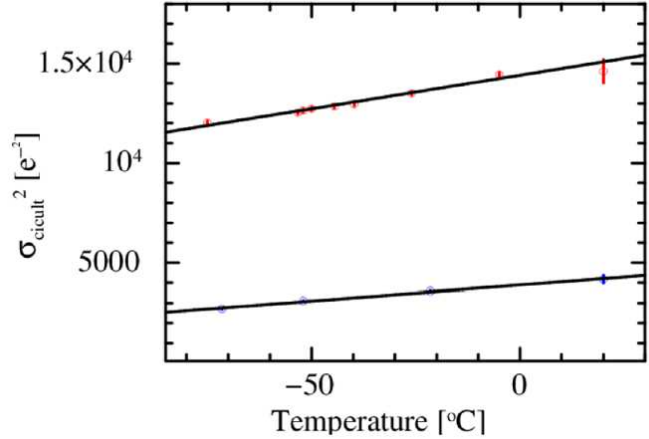}
\caption{The temperature dependence of the circuit origin noise. Data of XRPIX1(red) and XRPIX2b\_CZ (blue). Straight line is a linear function fitting results.}
\label{fig:circut_1}
\end{figure}

Figure \ref{fig:circut_1} shows the temperature dependence of the circuit origin noise.

In Figure \ref{fig:circut_1}, it was found that circuit origin noise is proportional to the square root of operating temperature.
Table \ref{tab:fit} shows the ftting results.

\begin{table}[H]
\begin{center}
\caption{Parameters of the model fit.}
\begin{tabular}{ccc} \hline \hline
& slope& intercept\\ \hline
XRPIX1& 31.5$ \pm$ 5.0& (1.43$\pm$0.03) $\times$ 10$^4$\\
XRPIX2b\_CZ& 15.8$ \pm$ 0.2& (3.89$\pm$0.07) $\times$ 10$^3$\\ \hline
\end{tabular}
\label{tab:fit}
\end{center}
\end{table}

\section{Summary}
We examined the temperature dependence of the gain, energy resolution, readout noise and leak current of XRPIX.
We found that the readout noise, energy resolution, and leak current is reduced by lowering the operating temperature.
Gain of XRPIX1 was constant regardless of temperature, and the gain of XRPIX2b\_CZ showed a temperature dependence.
By lowering the operating temperature, the Energy resolution, the readout noise and the leak current are 1.08 keV FWHM @ 5.95 keV, 6.76 e$^-$/ms/pixel, 110 e$^-$ in XRPIX1 or 
630 eV FWHM @ 5.95 keV, 41.6 e$^-$/ms/pixel, 52 e$^-$ in XRPIX2b\_CZ. 
In addition, Readout noise is isolated the leak current origin noise and the circuit origin noise and we have evaluated the circuit origin noise.
We found that the ecircuit origin noise was  proportional to the square root of operating temperature.

\Acknowledgements
The authors are grateful to all the members of the SOI group and the research cooperation extended by Lapis Semiconductor Co. Ltd.
This work is supported  by the Japan Socitey for the Promotion of Science (JSPS) KAKENHI (25109004).

\end{document}